\documentstyle[aps]{revtex}
\draft
\begin{document}
\title
{Invariant closures for  the Fokker--Planck equation
} 
\author{ 
Iliya V.\ Karlin\cite{IVK}\\
Computing Center RAS, Krasnoyarsk 660036,  Russia\\
V.\ B.\ Zmievskii\cite{VZ}\\
LMF/DGM, Swiss Federal Institute of Technology,
CH-1015 Lausanne, Switzerland
}
\maketitle
\begin{abstract}
We develop the principle of dynamic invariance to obtain closed
moment equations from the Fokker--Planck kinetic equation.
The analysis is carried out to  explicit formulae 
for computation of the lowest eigenvalue and of the corresponding
eigenfunction for arbitrary potentials.  
\end{abstract}
\pacs{05.20.Dd, 05.70.Ln, 83.10.-y}  
The Fokker--Planck equation (FPE) is a familiar model in various
problems of nonequilibrium statistical physics
\cite{VanKampen}. In this paper we
consider the FPE of the form  
\begin{equation}
\label{FP}
\partial_t W
=\partial_x
\cdot\!\left\{D\cdot\left[W\partial_x U
+\partial_x W\right]\right\}.
\end{equation}
Here $W(x,t)$ is the probability density over the configuration
space $x$, at the time $t$, while $U(x)$ and $D(x)$ are
the potential and the positively semi-definite ($ y\cdot
D\cdot y\ge
0$) diffusion matrix. The
dot denotes convolution in the configuration space.
The FPE (\ref{FP}) is particularly important in studies of 
polymer solutions \cite{Bird}.
Let us recall the two properties of the FPE (\ref{FP}), important to
what will follow:
(i). Conservation of the total probability: $
\int\!W(x,t)d
x=1.$
(ii). Dissipation: The equilibrium distribution,
$W_{eq}\propto\exp(-U)$, is the unique
stationary solution to the FPE (\ref{FP}). The entropy,
\begin{equation}
\label{entropy}
S[W]=-\int\!W(x,t)
\ln\left[\frac{W(x,t)}{W_{eq}(x)}\right]dx,
\end{equation}
is a monotonically growing function due to the FPE (\ref{FP}), and it
arrives at the global maximum in the equilibrium.
These
properties are most apparent when the FPE (\ref{FP}) is rewritten as follows:
\begin{equation}
\label{GENERIC}
\partial_{t}W(x,t)=\hat{M}_W\frac{\delta S[W]}{\delta W(x,t)},
\end{equation} 
where $\hat{M}_W=-\partial_x\cdot[W(x,t)D(x)\cdot\partial_x]$ is
a positive semi--definite symmetric operator with kernel $1$. 
The form (\ref{GENERIC})  
(the dissipative vector field is a metric
transform of the entropy gradient) is an example of the dissipative part of a
structure termed GENERIC in a recent series of papers  
\cite{GENERIC}. 

Usually one is interested in dynamics of moments of the distribution
function $W$ rather than in the dynamics of the $W$ itself. 
Except for simplest
potentials $U$ and diffusion matrices $D$, the moment equations, as
they follow from the FPE (\ref{FP}), are not closed. Therefore, closure
procedures are required.

In this paper we address the problem of closure for the FPE (\ref{FP})
in a general setting. 
First, we review the maximum
entropy principle (MEP) as a source of suitable initial approximations
for the closures. 
We also discuss a version
of the MEP, valid for a near--equilibrium dynamics, and which 
results in explicit formulae for arbitrary $U$ and $D$. 

The MEP closures are almost never invariants of the true moment dynamics,
and corrections to the MEP closures is the central issue of this
paper.
For this purpose, we apply the method of invariant manifold \cite{KAM},
which is carried out (subject to certain approximations explained below) 
to explicit recurrence formulae for 
one--moment near--equilibrium closures for arbitrary $U$ and $D$.
These formulae give a method for computing the lowest eigenvalue 
of the problem, and which dominates the near--equilibrium FPE dynamics.

{\it MEP closures}\cite{Comment1}. 
Let $M=\{M_0,M_1,\dots,M_k\}$ be linearly independent moments
of interest, $M_i[W]=\int m_i(x)W(x)dx$, and where $m_0=1$. We
assume existence a function $W^*(M,x)$ which extremizes the entropy
$S$ (\ref{entropy}) under the constraints of fixed $M$. This 
MEP distribution function may be written
\[
W^*=W_{eq}\exp\left[\sum_{i=0}^{k}\Lambda_im_i(x)-1\right],
\]
where $\Lambda=\{\Lambda_0,\Lambda_1,\dots,\Lambda_k\}$ are Lagrange
multipliers. Closed equations for moments $M$ are derived in two steps.
First, the MEP distribution  is substituted into the FPE (\ref{FP}) 
or (\ref{GENERIC}) to give a formal expression:
$\partial_t W^*=\hat{M}_{W^*}(\delta S/\delta W)\big|_{W=W^*}$. 
Second, applying a projector $\Pi^*$,
\[
\Pi^*\bullet=\sum_{i=0}^{k}(\partial W^{*}/\partial M_i)\int m_i(x)\bullet
dx,
\]
on both sides of this formal expression,
we derive closed equations for $M$ in the MEP approximation. Further 
processing requires an explicit solution to the constraints,
$\int W^*(\Lambda,x)m_i(x)dx=M_i$, to get
the dependence of Lagrange multipliers $\Lambda$ on the moments $M$. 
Though typically the
functions $\Lambda(M)$ are not known explicitly,
one general remark about the moment equations is readily available.
Specifically, the moment equations in the MEP approximation have
the form:
\begin{equation}
\label{GENERIC2}
\dot{M}_i=\sum_{j=0}^k M^*_{ij}(M)\frac{\partial S^*(M)} 
{\partial M_j},
\end{equation}
where $S^*(M)=S[W^*(M)]$ is the macroscopic
entropy, and where $M^*_{ij}$ is an $M$-dependent $(k+1)\times(k+1)$ matrix:
\[
M^*_{ij}=\int W^*(M,x)[\partial_x m_i(x)]\cdot D(x)\cdot
[\partial_x m_j(x)] dx.
\]
The matrix $M^*_{ij}$ is symmetric, 
positive semi--definite, and its kernel is the vector $\delta_{0i}$. 
Thus,  
{\it the MEP closure reproduces the GENERIC structure
on the macroscopic level}, the vector field of macroscopic
equations (\ref{GENERIC2}) is a metric transform of the gradient
of the macroscopic entropy. 

{\it Triangle MEP closures} \cite{Comment2}.  
The following
version of the MEP makes it possible to derive more explicit results 
in a general setting: 
In many cases, one can split the set of moments $M$ in two parts,
$M_{I}=\{M_0,M_1,\dots,M_l\}$ and $M_{II}=\{M_{l+1},\dots,M_k\}$, in
such a way that the MEP distribution can be constructed explicitly for
$M_{I}$ as $W_{I}^*(M_{I},x)$. 
The full MEP problem for $M=\{M_{I},M_{II}\}$
in the "shifted" formulation reads:
extremize the functional $S[W_I^*+\Delta W]$ with respect to 
$\Delta W$, subject to the constraints 
$M_{I}[W_I^*+\Delta W]=M_{I}$ and $M_{II}[W_I^*+\Delta W]=M_{II}$.
Let us denote as $\Delta M_{II}$ 
deviations of the moments $M_{II}$ from their values in the 
state $W_I^*$. For small deviations, the entropy 
is well approximated with a quadratic functional $\Delta S[\Delta W]$
which is an expansion of the functional (\ref{entropy}) in the state
$W^*_I$ up to the terms of the order $\Delta W^2$.
With $M_I[W^*_I]=M_I$, we come to the following
problem: extremize the functional
$\Delta S[\Delta W]$, subject to the constraints $M_I[\Delta W]=0$,
and $M_{II}[\Delta W]=\Delta M_{II}$.
The solution to the latter problem is always explicitly found
from a $(k+1)\times(k+1)$ system of linear algebraic equations 
for Lagrange multipliers. 

In the remainder of this paper we deal with one--moment
near--equilibrium closures:  
$M_{I}=M_0$, (i.\ e. 
$W_I^*=W_{eq}$), and the set $M_{II}$
contains a single moment $M=\int mWdx$, $m(x)\neq 1$.
We will specify
notations for the near--equilibrium FPE , writing the distribution
function as $W=W_{eq}(1+\Psi)$, where the function $\Psi$
satisfies an equation:
\begin{equation}
\label{FPLIN}
\partial_t \Psi=W_{eq}^{-1}\hat{J}\Psi,
\end{equation}
where $\hat{J}=\partial_x\cdot[W_{eq}D\cdot\partial_x]$.
The triangle one--moment  
MEP function reads: 
\begin{equation}
\label{initial}
W^{(0)}=W_{eq}\left[1+\Delta M m^{(0)}\right]
\end{equation}
where $\Delta M=M-\langle m\rangle$, and 
\begin{equation}
\label{INDIR}
m^{(0)}=
[\langle m m\rangle-\langle m \rangle^2]^{-1}
[m-\langle m \rangle].
\end{equation}
Brackets $\langle\dots\rangle=\int W_{eq}\dots dx$ 
denote equilibrium averaging.
The superscript $(0)$ indicates
that the triangle MEP function (\ref{initial}) will be considered
as an initial approximation to a procedure which we address below.
Projector for the approximation (\ref{initial})
has the form
\begin{equation}
\label{P0}
\Pi^{(0)}\bullet=W_{eq}\frac{m^{(0)}}
{\langle m^{(0)}m^{(0)}\rangle}\int  m^{(0)}\bullet dx.
\end{equation}
substituting the 
function (\ref{initial}) into the FPE (\ref{FPLIN}), and applying the
projector (\ref{P0}) on both the sides of the resulting formal
expression, we derive an equation for $M$:
$\dot{M}=-\lambda_0\Delta M$,
where $1/\lambda_0$ is 
the inverse effective time of relaxation of the moment $M$ to its equilibrium
value, in the MEP approximation (\ref{initial}):                
\begin{equation}
\label{TIME0}
\lambda_0=\langle m^{(0)}m^{(0)} \rangle^{-1}
\langle \partial_x m^{(0)}\cdot D \cdot\partial_x m^{(0)}\rangle.
\end{equation} 
  
{\it Invariant closures.} Both the MEP and
the triangle MEP closures are
almost never invariants of the FPE dynamics. That is, the moments $M$ of
solutions to the FPE (\ref{FP}) vary in time differently from the solutions
to the closed moment equations like (\ref{GENERIC2}), and these variations
are generally significant even for the near--equilibrium dynamics.  
Therefore, we ask for corrections to the MEP closures to finish with
the invariant closures
\cite{KAM}.

First, the invariant one--moment closure is given by an unknown 
distribution function $W^{(\infty)}=W_{eq}[1+\Delta M m^{(\infty)}(x)]$
which satisfies an equation
\begin{equation}
[1-\Pi^{(\infty)}]\hat{J}m^{(\infty)}=0.
\label{INV}
\end{equation}
Here $\Pi^{(\infty)}$ is a projector, associated with an unknown function
$m^{(\infty)}$, and which is also yet unknown. Eq.\ (\ref{INV})
is a formal expression of the invariance principle for a one--moment
near--equilibrium closure: considering $W^{(\infty)}$ as a 
manifold in the space of distribution functions, parameterized with
the values of the moment $M$, we require that the microscopic 
vector field $\hat{J}m^{(\infty)}$ be equal to its projection,
$\Pi^{(\infty)}\hat{J}m^{(\infty)}$,
onto the tangent space of the manifold $W^{(\infty)}$.

Now we turn our attention to solving the invariance equation
(\ref{INV}) iteratively, beginning with the triangle one--moment
MEP approximation $W^{(0)}$ (\ref{initial}).
We apply the following iteration process to the Eq.\ (\ref{INV}):
\begin{equation}
[1-\Pi^{(k)}]\hat{J}m^{(k+1)}=0,
\label{ITERATIONS}
\end{equation}
where $k=0, 1,\dots$, and 
where $m^{(k+1)}=m^{(k)}+\mu^{(k+1)}$, and the correction
satisfies the condition $\langle \mu^{(k+1)}m^{(k)}\rangle=0$.
Projector is updated after each iteration,
and it has the form
\begin{equation}
\Pi^{(k+1)}\bullet=W_{eq}\frac{m^{(k+1)}}
{\langle m^{(k+1)}m^{(k+1)}\rangle}\int m^{(k+1)}(x)\bullet dx.
\label{Pk}
\end{equation}
Applying $\Pi^{(k+1)}$ to the formal expression, 
$W_{eq}m^{(k+1)}\dot{M}=\Delta M[1-\Pi^{(k+1)}]\hat{J}m^{(k+1)}$, 
we derive the macroscopic equation, $\dot{M}=-\lambda_{k+1}\Delta M$, where
$\lambda_{k+1}$ is the $(k+1)$th update of the 
inverse effective time (\ref{TIME0}): 
\begin{equation}
\label{LAMBDAk}
\lambda_{k+1}=
\frac{\langle \partial_x m^{(k+1)}\cdot D \cdot\partial_x m^{(k+1)}\rangle}
{\langle m^{(k+1)}m^{(k+1)}\rangle}.
\end{equation}
Specializing to the one--moment near--equilibrium closures, and 
following a general argument \cite{KAM}, solutions to the invariance equation
(\ref{INV}) are eigenfunctions of the operator $\hat{J}$, while
the formal limit of
the iteration process (\ref{ITERATIONS}) is the eigenfunction
which corresponds to the eigenvalue
with the minimal nonzero absolute value.

{\it Diagonal approximation.} 
To obtain more
explicit results in the iteration process (\ref{ITERATIONS}), 
we introduce an approximate solution
{\it on each iteration}. 
The correction $\mu^{(k+1)}$ satisfies the condition
$\langle m^{(k)}\mu^{(k+1)}\rangle=0$, and  
can be decomposed as follows:
$\mu^{(k+1)}=\alpha_k e^{(k)}+e^{(k)}_{ort}$.
Here 
$e^{(k)}=W_{eq}^{-1}[1-\Pi^{(k)}]\hat{J}m^{(k)}=\lambda_k m^{(k)}+R^{(k)}$
is the variance of the $k$th approximation,
where
\begin{equation}
R^{(k)}=\partial_x\cdot[D\cdot \partial_x m^{(k)}]
-\partial_x U\cdot D\cdot \partial_x m^{(k)}.
\label{Rk}
\end{equation}
The function $e^{(k)}_{ort}$ is orthogonal to both
$e^{(k)}$ and $m^{(k)}$: $\langle e^{(k)}e^{(k)}_{ort}\rangle=0$,
and $\langle m^{(k)}e^{(k)}_{ort}\rangle=0$. 
Our diagonal approximation (DA) consists in disregarding the part
$e^{(k)}_{ort}$. 
Specifically, we consider the following ansatz at the $k$th iteration: 
\begin{equation}
m^{(k+1)}=m^{(k)}+\alpha_k e^{(k)}.
\label{ANSATZ}
\end{equation} 
Substituting the ansatz (\ref{ANSATZ}) into the  
Eq.\ (\ref{ITERATIONS}),
and integrating the latter expression with the function
$e^{(k)}$, we evaluate the coefficient $\alpha_k$:

\begin{equation}
\alpha_k=\frac{A_k-\lambda_k^2}{\lambda_k^3-2\lambda_k A_k+B_k},
\label{ALPHA}
\end{equation}
where parameters $A_k$ and $B_k$ represent the following equilibrium
averages: 
\begin{eqnarray}
\label{CONSTANTS}
A_k&=&\langle m^{(k)} m^{(k)}\rangle^{-1}
\langle R^{(k)}R^{(k)}\rangle\\\nonumber
B_k&=&\langle m^{(k)} m^{(k)}\rangle^{-1}
\langle \partial_x R^{(k)}\cdot D\cdot\partial_x R^{(k)}\rangle.
\end{eqnarray}

Finally, putting together Eqs.\ (\ref{LAMBDAk}), (\ref{Rk}),
(\ref{ANSATZ}), (\ref{ALPHA}), and (\ref{CONSTANTS}), we 
arrive at the following DA recurrency solution, and which is our main
result:
\begin{mathletters}
\label{RESULT}
\begin{eqnarray}
m^{(k+1)}&=&m^{(k)}+\alpha_k[\lambda_k m^{(k)}+R^{(k)}],\\
\lambda_{k+1}&=&\frac{\lambda_k-(A_k-\lambda_k^2)\alpha_k}
{1+(A_k-\lambda_k^2)\alpha_k^2}.
\label{LAMBDA}
\end{eqnarray}
\end{mathletters} 

To test the convergency of the DA process (\ref{RESULT})
we have considered two potentials $U$ in the FPE (\ref{FP}) 
with a constant diffusion matrix $D$. The first test was with 
the square potential $U=x^2/2$, in the three--dimensional configuration
space, since for this potential the detail structure of the spectrum is
well known.   
We have considered two examples of initial one--moment MEP closures
with $m^{(0)}=x_1+100(x^2-3)$ (example 1), and
$m^{(0)}=x_1+100x^6 x_2$ (example 2),
in the Eq.\ (\ref{INDIR}). The result of performance of the DA for
$\lambda_k$ (\ref{LAMBDA})
is presented in the Table \ref{Tab1}, together with the error $\delta_k$ 
which was
estimated as the norm of the variance at each iteration:
$\delta_k=
\langle e^{(k)}e^{(k)}\rangle/\langle m^{(k)}m^{(k)}\rangle$.   
In both examples, we see a good monotonic convergency to the 
minimal eigenvalue $\lambda_{\infty}=1$, corresponding to the eigenfunction
$x_1$. This convergency
is even striking in the
example 1, where the initial choice was very close to a different eigenfunction
$x^2-3$, and which can be seen in the non--monotonic behavior of the
variance. Thus, we have an example to trust the DA 
as converging to the stationary point of the original iteration procedure
(\ref{ITERATIONS}).

For the second test, we have taken a one--dimensional potential
$U=-50\ln(1-x^2)$, the configuration space is the segment
$|x|\leq 1$. Potentials of this type (so--called FENE potential)
are used in applications of the FPE to models of polymer solutions
\cite{Bird}. Results are given in the Table \ref{Tab2} for
the two initial functions, 
$m^{(0)}=x^2+10x^4-\langle x^2+10x^4 \rangle$ (example 3),
and
$m^{(0)}=x^2+10x^8-\langle x^2+10x^8 \rangle$ (example 4).
Both the examples demonstrate a stabilization of the $\lambda_k$
at the same value after some ten iterations. 

In conclusion, we have developed the principle of invariance to
obtain moment closures for the Fokker--Planck equation (\ref{FP}),
and have derived explicit results for the one--moment near--equilibrium
closures, particularly important to get information about the
spectrum of the FP operator.

I.\ V.\ K. is thankful to M.\ Grmela for providing a preprint
of their paper \cite{GENERIC}, and to H.\ C.\ \"{O}ttinger for
stimulating discussions. V.\ B.\ Z. is thankful to J.--C.\ Badoux 
and M.\ Deville for the possibility of the research stay at the
EPFL. The work of I.\ V.\ K. was supported by 
CNR, and by RFBR through grant No.\ 95-02-03836-a.

\begin{table}[p]
\caption{
Iterations $\lambda_k$ and the error  
$\delta_k$
for $U=x^2/2$.}  
\begin{tabular}{|c|c|c|c|c|c|c|c|c|}
\hline &
& $0$ & $1$ & $4$ & $8$ 
& $12$ & $16$  &  $20$ \\  
\cline{2-9}  Ex.\ 1& $\lambda$  
& 1.99998 & 1.99993 & 1.99575 
& 1.47795 & 1.00356 & 1.00001 & 1.00000 
\\\cline{2-9} & $\delta $ 
& $0.16 \cdot 10^{-4}$ & $ 0.66 \cdot 10^{-4}$ & $ 0.42 \cdot 10^{-2}$ & $0.24$ 
& $0.35 \cdot 10^{-2}$ & $ 0.13 \cdot 10^{-4}$ & $ 0.54 \cdot 10^{-7}$ \\ 
\hline 
& & $ 0$ & $1$ & $2$ & $3$ 
& $ 4$ & $ 5$  &  $6$ 
\\  \cline{2-9} Ex.\ 2 &  $\lambda$ 
& 3.399 & 2.437 & 1.586 & 1.088 & 1.010 
& 1.001 & 1.0002 \\ 
\cline{2-9} & $\delta$ 
& $1.99$ & $1.42$ & $0.83$ &  $0.16$ & $0.29 \cdot 10^{-1}$ & $0.27 \cdot 10^{-2}$ 
& $0.57 \cdot 10^{-3}$ \\
 \hline
\end{tabular}
\label{Tab1}
\end{table}

\begin{table}[p]
\caption{Iterations $\lambda_k$   
for  
$U=-50\ln(1-x^2)$. 
}  
\begin{tabular}{|c|c|c|c|c|c|c|c|c|c|c|}
\hline 
 & &
 $0$ & $1$ & $2$ & $3$ & 
$ 4$ & $ 5$  &  $6$ &$7$& $8$ \\  
\hline 
Ex.\ 3 & 
 $\lambda$ &
213.1774 & 
 212.1864 & 211.9148 & 211.8619 & 211.8499 & 211.8453 & 211.8433 &
 211.8422 & 211.8417 
\\ \hline 
Ex.\ 4 &
 $\lambda$ &
 216.5856 & 213.1350 & 212.2123 & 211.9984 & 
211.9295 & 211.8989 & 211.8838 & 211.8757 & 211.8713 
 \\ \hline 
\end{tabular}
\label{Tab2}
\end{table}

\end{document}